# Tight lower bounds for online labeling problem

Jan Bulánek[*]    Michal Koucký[†]    Michael Saks[‡]

August 30, 2018

**Abstract**

We consider the *file maintenance problem* (also called the *online labeling problem*) in which $n$ integer items from the set $\{1, \ldots, r\}$ are to be stored in an array of size $m \geq n$. The items are presented sequentially in an arbitrary order, and must be stored in the array in sorted order (but not necessarily in consecutive locations in the array). Each new item must be stored in the array before the next item is received. If $r \leq m$ then we can simply store item $j$ in location $j$ but if $r > m$ then we may have to shift the location of stored items to make space for a newly arrived item. The algorithm is charged each time an item is stored in the array, or moved to a new location. The goal is to minimize the total number of such moves the algorithm has to do. This problem is non-trivial when $n \leq m < r$.

In the case that $m = Cn$ for some $C > 1$, algorithms for this problem with cost $O(\log(n)^2)$ per item have been given [IKR81, Wil92, BCD+02]. When $m = n$, algorithms with cost $O(\log(n)^3)$ per item were given [Zha93, BS07]. In this paper we prove lower bounds that show that these algorithms are optimal, up to constant factors. Previously, the only lower bound known for this range of parameters was a lower bound of $\Omega(\log(n)^2)$ for the restricted class of *smooth* algorithms [DSZ05a, Zha93].

We also provide an algorithm for the sparse case: If the number of items is polylogarithmic in the array size then the problem can be solved in amortized constant time per item.

## 1 Introduction

**The file maintenance problem**

In this paper we consider the *file maintenance problem* in which $n$ integer items from the set $\{1, \ldots, r\}$ are to be stored in an array of size $m \geq n$. The items are presented sequentially in an arbitrary order, and must be stored in the array in sorted order (but not necessarily in consecutive locations). Each new item must be stored in the array before the next item is received. If $r \leq m$ then we can simply store item $j$ in location $j$ but if $r > m$ then we may have to shift the location of stored items to make space for a newly arrived item. The algorithm is charged each time an item is stored in the array, or moved to a new location. The goal is to minimize the total number of such moves the algorithm has to do. This problem is non-trivial when $n \leq m < r$.

[*]MFF UK and Institute of Mathematics, Academy of Sciences, Prague, e-mail: honyai@seznam.cz. Partially supported by GAUK project no. 344711, grant GA ČR P202/10/0854, project No. 1M0021620808 of MŠMT ČR, Institutional Research Plan No. AV0Z10190503 and grant IAA100190902 of GA AV ČR.

[†]Institute of Mathematics, Academy of Sciences, Prague, e-mail: koucky@math.cas.cz. Currently on sabbatical at the University of Toronto partially supported by NSERC. Partially supported by GA ČR P202/10/0854, project No. 1M0021620808 of MŠMT ČR, Institutional Research Plan No. AV0Z10190503 and grant IAA100190902 of GA AV ČR.

[‡]Department of Mathematics, Rutgers University, e-mail: saks@math.rutgers.edu. Supported in part by NSF under grant CCF-0832787.

An alternate formulation is the *online labeling problem* in which arriving items must be assigned an integer label in the range $[1, m]$ so that the order on the labels agrees with the numerical ordering on the items. The algorithm pays one each time an item is labeled or relabeled. Typically in the literature the file maintenance problem refers to the *small space regime* in which $m = O(n)$. This case is the focus of this paper.

Itai et al. [IKR81] were the first to design an algorithm that maintains an array of size $m = O(n)$ sorted while making only $O(n \log(n)^2)$ moves in total, i.e., in amortized setting the algorithm makes $O(\log(n)^2)$ moves per item. Willard [Wil92] improved this algorithm to the worst case setting of $O(\log(n)^2)$ moves per item and Bender et al. [BCD+02] further simplified his result. The Itai et al. approach can be modified so that for an array of size $m = n^{1+\epsilon}$, $\epsilon > 0$ constant, it uses only $O(\log(n))$ moves per item, amortized (folklore). For the case that the array size $m$ is exactly the number of items $n$, [Zha93] gave an algorithm that achieves a surprising amortized bound $O(\log(n)^3)$ moves per item; this result was simplified in [BS07].

In recent years there has been renewed interest in this problem due to its applications in the design of cache-oblivious algorithms, e.g., design of cache-oblivious B-trees [BDFC05, BFJ02] and cache-oblivious dynamic dictionaries [BDIW04]. However, until now it was not known whether the maintenance algorithms for the small space regime can be improved to achieve better amortized cost.

## Our results

In this paper we prove an $\Omega(n \log(n)^2)$ lower bound on the number of moves for inserting $n$ items into array of size $m = O(n)$ for any online labeling algorithm, matching the known upper bound up to constant factors. For the case of array size $m = n + n^{1-\epsilon}$, we prove the asymptotically optimal lower bound $\Omega(n \log(n)^3)$.

Our lower bounds are valid even for relatively small $r$; it is enough that $r$ is bounded below by a sufficiently large constant times $m$. (Recall that the problem has a trivial solution of cost $n$ if $r \leq m$.)

Our lower bounds apply to slightly superlinear array size. For example, if $m = O(n \log(n)^{1-\varepsilon})$ one can prove an amortized lower bound $\Omega(\log(n)^{1+\varepsilon/3})$ (though here we need a large range size to get this lower bound.)

Our bound is the first lower bound for general algorithms in the small space regime. Previously Dietz et al. [DSZ05a, Zha93] proved an amortized $\Omega(\log(n)^2)$ lower bound for the restricted case of so-called *smooth algorithms*.

In addition to the lower bounds, we provide a new upper bound in the case than $m$ is a large function of $n$. We give an algorithm that provided that $m$ is at least $2^{\log(n)^k}$ for $k \geq 3$ has amortized cost $O(\log(n)/\log\log(m))$. In particular, for any fixed $c$, $\log(m)^c$ items can be inserted into an array of size $m$ in constant amortized time.

## Previous lower bounds

There are two previous papers, both by Dietz, Seiferas and Zhang, that give lower bounds for this problem. The first ([DSZ05a], also available in Zhang's Ph.D. thesis [Zha93]) considers the small space regime, and proved an $\Omega(\log(n)^2)$ amortized lower bound for a restricted class of algorithms, called *smooth algorithms*, which are limited to redistributing items in a uniform fashion. While this lower bound is interesting and non-trivial (and introduces several key ideas that we use in our lower bound), the restriction to smooth algorithms is significant. The lower bound for smooth algorithms is obtained by considering a very simple adversary which exploits the smoothness of the algorithm; a non-smooth algorithm can easily handle the given adversary with constant amortized time per item as the adversary inserts all the items in decreasing order. There is some confusion in the literature about this result, the fact that it applies only to a restricted class of algorithms is sometimes not mentioned (e.g., [BS07]), creating the impression that the general lower bound result was already established.

The other lower bound for this problem is the amortized $\Omega(\log(n))$ lower bound of Dietz et al. [DSZ05b] for arrays of size $m = n^{O(1)}$. His lower bound applies to the "intermediate space" regime (polynomial in the number of items), which is not dealt with in this paper. We have some concerns about the correctness of this paper which we have recently raised with one of the authors (Seiferas), who agrees that there is an error that does not seem to be readily fixable. (Their result consists of two parts: a lower bound for a problem they call bucketing which we believe is correct, and a reduction which converts the lower bound for bucketing to a lower bound for online labeling, which we have doubts about.)

Whether or not this reduction turns out to be valid, this paper, like the other paper mentioned above, lays out the basic approach and provides important ideas which we make use of in our paper.

## Proof technique

The general idea for our lower bound (which builds heavily on the above-mentioned work of Dietz et al.) is to build an adversary that will force the maintenance algorithm to make many moves of items that are already stored in the array. The adversary will attempt to identify a densely populated (*crowded*) segment of the array and load an item that is in the middle of the items already stored there. Repeated insertions of items with value in this range will eventually force the algorithm to move the existing items.

Deriving a lower bound based on this idea has various complications. The natural notion of crowding of a segment is the ratio of stored items to the size of the segment. Whether a particular portion of the array is considered to be crowded may depend on the scale of segments being considered; there may be a relatively small segment that is very crowded, but larger segments containing it are uncrowded. To force the algorithm to work hard, we want to identify a region that is crowded at many different scales. This suggests identifying a long nested sequence of segments covering a wide range of scales, such that each is crowded. The hope is that loading many items having value in the middle of the range of items stored in the smallest nested segment will force the algorithm to do costly rearrangements at all of these different scales.

A straightforward way to accomplish this is to start with the entire array, and successively select a nested subsegment having highest density among subsegments of, say, half the size of the current segment. This results in a sequence of segments of increasing density, but does not seem to be enough to give a good lower bound. The problem arises when successive selected subsegments are chosen near the boundary of the parent segment. In this case, the algorithm may be able to relieve overcrowding by relatively inexpensive rearrangements that cross the boundary of many segments in the sequence into uncrowded segments. To avoid this, the adversary would like to select each subsegment in the sequence so that it has a significant buffer to its left and right in the parent segment, where each buffer contains a constant fraction of the items in the parent segment. The presence of such buffers can be used to ensure that as a segment gets crowded, all of the items in either its left or right buffer will have to be moved. The difficulty is that when we insist on having these buffers we can no longer ensure that the density of the segments in the sequence do not decrease (because a given segment in the sequence may have its items concentrated near its boundary). So we have to allow some decrease in the segment density along the sequence.

Dietz et al. [DSZ05b] manage to construct such a nested sequence in which each successive segment has a large left and right buffer. The problem is that the density of segments down the sequence may decrease by as much as a constant factor, so that if the sequence has logarithmic length the density decreases by a fraction $n^{\Omega(1)}$. This limits the quality of lower bounds that can be proved.

The goal then is to define this nested sequence in such a way that we still have large buffers, but the density degrades at a much slower rate. Our approach begins with the observation that if for a given segment every subsegment having large buffers has density significantly smaller than the given segment, then there must be a large subsegment (near the boundary of the given segment) that does not have large buffers but does has substantially higher density than the given segment. This allows us to build a chain of $\Theta(\log(n))$ segments, such that most of the segments have large buffers with respect to their parents, and the degradation of density along the entire chain can be bounded by a constant factor. (To give a rough idea of the choice of parameters, when $m = \Theta(n)$, we allow decrease in density by a factor of at most $(1 - \Theta(1/\log(n)))$ or increase by a factor of at least $(1 + \Theta(1/\log(n)))$ in a single step.)

After identifying such a sequence of nested segments the adversary inserts new items into the inner most segment. Whenever the maintenance algorithm rearranges some portion of the array the adversary rebuilds the affected portion of the segment chain. An accounting similar to that of Dietz at el. [DSZ05a, Zha93] can then be applied on the segments having large buffers, to obtain the lower bound $\Omega(\log(n)^2)$.

In our actual implementation of this idea we don't explicitly deal with the high density unbuffered segments. Rather we construct the sequence of segments in such a way that each segment in the sequence satisfies a strong uniformity property: No subsegment of the given segment of size at least $1/4$ of the given segment has density significantly smaller than the given segment. In searching for the successor segment of a given segment $S$ with this uniformity property, we first restrict to the middle third $T$ of the segment, which we are guaranteed has density close to that of $S$. We then look for a large subsegment of $T$ having density close to that of $T$ and having the desired uniformity property. The restriction to $T$ means that when we choose the next segment it is guaranteed to have large buffers. To identify the desired subsegment of $T$ we maximize a certain potential function defined on the subsegments of $T$ that gives a large subsegment $D$ of $T$ that almost has the needed uniformity properties: we get the needed properties by taking the middle third of $D$, and this is the next segment in the sequence. Maximizing the potential function implicitly captures the process of successively choosing subsegments of significantly higher density until one arrives at a subsegment for which no such selection is possible.

The $\Omega(\log(n)^3)$ lower bound for an array of size $m = n$ is obtained by iterative application of the $\Omega(\log(n)^2)$ lower bound for inserting always one half of the remaining items. This parallels the idea used in [Zha93] to obtain a matching upper bound.

For the lower bounds, our adversary needs some room in the range of values to select new items that should be inserted into the array. Once there are two keys in the array that are consecutive elements in the range of values the adversary cannot choose another element to be stored in between them in the array. As he inserts more items into the same position in the array the available room shrinks. This limits his power. To mitigate this problem we assign a slightly smaller weight to newer items that are inserted. Since our adversary tries to select a sequence of dense nested segments it automatically avoids places crowded by newer items. This technique allows us to bound the range size.

## 2 The model and main result

### 2.1 A two player game

In this paper, interval notation is used for sets of consecutive integers, e.g., $[a, b]$ is the set $\{k \in \mathbb{Z} : a \leq k \leq b\}$. We consider an array with cells indexed by the set $[1, m]$ in which we store a set $Y$ of integer-valued *keys*. A *storage function* for $Y$ is a map $f : Y \longrightarrow [1, m]$ that is strictly order preserving, i.e., for $x, y \in Y$ if $x < y$ then $f(x) < f(y)$. In particular $f$ is one-to-one, so $|Y| \leq m$. Cells that are in the image of the map $f$ are said to be *occupied* and the others are said to be *unoccupied*. A *configuration* is a pair $(Y, f)$ where $Y$ is a set of keys and $f$ is a storage function for $Y$.

To formalize the array loading problem we define a game $G^n(m, r)$, where $n, m, r$ are positive integer parameters, which is played by two players, the *adversary* and the *algorithm*. The game is played in a sequence of $n$ time steps. At step $t$, the *adversary* selects a key $y^t$ from the set $\{1, \ldots, r\} - \{y^1, \ldots, y^{t-1}\}$, and the algorithm responds with a storage function $f^t$ for the set $Y^t = \{y^1, \ldots, y^t\}$. We say that key $y^t$ is *loaded* at step $t$. $(Y^t, f^t)$ is called the *configuration at step $t$*.

A key $y$ is *relocated* at step $t$ if $f^t(y) \neq f^{t-1}(y)$. In particular $y^t$ is relocated at step $t$. The set of relocated keys at step $t$ is denoted $Rel^t$. The *cost up to step $t$* is $\chi^t = \sum_{i=1}^{t} |Rel^i|$. Clearly $\chi^t \geq t$ for every $t$. The objective of the algorithm is to minimize $\chi^n$ and the objective of the adversary is to maximize $\chi^n$. We write $\chi^n(m, r)$ for the smallest cost that can be achieved by the algorithm against the best adversary.

$G^n(m, r)$ is not well defined if $n > m$ since there can be no storage function once the number of items exceeds the number of cells. Also, if $m \geq r$, there is a trivial algorithm that achieves optimal cost $n$ by storing each key $y \in [r]$ in cell $y$. We therefore assume $n \leq m < r$.

### 2.2 The main theorems

In this section, we state our lower bound results for $\chi^n(m, r)$. We divide our results into two theorems, corresponding to the relative size of $m$ and $n$.

The first theorem applies whenever $2n \leq m$, but it only gives interesting results provided that $m$ is not too large (slightly superlinear function of $n$). Here we separately consider two cases. In the first case, the range of possible keys is exponential in $n$. In the second case the range of keys is limited to be a constant times $m$. Despite this strong limitation, the lower bound is only slightly worse.

**Theorem 1** *There is a constant $C_1$ so that the following holds. Let $m, n$ be integers satisfying $C_1 \leq n$ and $2n \leq m$. Let $\delta = n/m$. Then*

1. *If $r \geq n2^{n-1}$ then $\chi^n(m, r) \geq n(\ln(n))^2 \frac{\delta}{C_1(\ln(1/\delta))^2}$.*

2. *If $r \geq C_1 m$ then $\chi^n(m, r) \geq n(\ln(n))^2 \frac{\delta^2}{C_1(\ln(1/\delta))^2}$.*

In both parts, if $m = O(n)$ then the lower bound obtained is $\Omega(n(\ln(n))^2)$. The first bound gives a nontrivial result (larger than $n$) for $m$ up to $\Theta(n \ln(n)^2/(\ln \ln(n))^2)$, while the second bound is nontrivial for $m$ up to $\Theta(n \ln(n)/\ln \ln(n))$.

In the next result we consider the $n \leq m < 2n$:

**Theorem 2** *There are constants $C_0, C_2, C_3$ so that the following holds. Let $m, n$ be integers satisfying $C_0 \leq n < m < 2n$ and let $\delta = n/m$. Assume $r \geq (\frac{1}{1-\delta})^{C_2} n$. Then:*

$$\chi^n(m,r) \geq \frac{1}{C_3} n (\ln(n))^2 \ln(1/(1-\delta)). \tag{1}$$

For $m \leq n + n^{1-\varepsilon}$ this gives a tight lower bound of $\Omega(n(\ln(n))^3)$. Observe that for this lower bound we only need the range of keys to be polynomial in $m$. (A more refined analysis can provide an asymptotically same lower bound with range size $n + O(n^{1-\varepsilon})$ for this case.)

## 2.3 Partially loaded arrays and the main lemma

As the game has been defined, every cell is initially unoccupied. For the proofs of the main theorems, it will be convenient to consider a small variant of the game, in which the array is initially partially loaded. This version of the game is specified by the parameters $n, m$ (but not $r$) and additionally takes a set $Y^0$ of keys, whose size is denoted by $n_0$. The array is initially loaded with the subset $Y^0$ and the algorithm selects the initial storage function $f^0$ (at no cost). The game then proceeds as before, except that the adversary is restricted to loading keys in the range $(\min(Y^0), \max(Y^0))$. We denote the game by $G^n(m|Y^0)$ and write $\chi^n(m|Y^0)$ for the minimum cost that can be achieved by the algorithm against the best adversary. We assume that $m \geq n_0 + n$, otherwise there is not enough room to load all of the keys.

For a set $Y$ of keys, we define mingap($Y$) to be the minimum absolute difference between pairs of keys in $Y$. As we will see, the following lemma easily implies Theorems 1 and 2.

**Lemma 3** *There are positive constants $C_0, C_4$ so that the following holds. Let $m, n, n_0$ be integers satisfying $C_0 \leq n \leq n_0$ and $n + n_0 \leq m$. Let $\delta_0 = n_0/m$. Assume $\delta_0 \in (\ln(n)^{-2}, 1 - n^{-1/5})$.*

*Let $Y^0$ be any set of $n_0$ keys. Let $\mu_0 = \text{mingap}(Y_0)$. Assume $\mu_0 \geq 1 + 12/\delta_0$.*

1. *If $\mu_0 \geq 2^n$ then $\chi^n(m|Y) \geq n(\ln(n))^2 \frac{\delta_0(1-\delta_0)}{C_4(\ln(1/\delta_0))^2}$.*

2. *If $\mu_0 < 2^n$, then $\chi^n(m|Y) \geq n(\ln(n))^2 \frac{\delta_0^2(1-\delta_0)}{C_4(\ln(1/\delta_0))^2}$.*

**Remark.** In the second part of the lemma, we assume only that $\mu_0 \geq 1 + 12/\delta_0$. If $\mu_0$ is much larger than $1 + 12/\delta_0$ we can do a simple "black box" modification of the adversary so that an additional property holds: At the conclusion of the game mingap$(Y^n) \geq \lfloor \mu_0/\lceil 1 + 12/\delta_0 \rceil \rfloor$. Here is how we modify the adversary. For each pair of keys in $Y^0$ that are adjacent (no intervening key in $Y^0$) the adversary selects $\lceil 12/\delta_0 \rceil$ equally spaced keys and ignores all other keys. Only these selected keys will be loaded during the game, so effectively the adversary is working with a mingap of $\lceil 1 + 12/\delta_0 \rceil$. Thus Lemma 3 can be applied to this restricted set of keys and at completion, the mingap is at least $\lfloor r/\lceil 1 + 12\delta_0 \rceil \rfloor$.

In the proofs of Theorems 1 and 2 we apply Lemma 3. When we apply the lemma, the parameter $n$ that appears in the theorem will not be the same as the parameter $n$ that appears in the lemma (but the meaning of the parameter $m$ does not change.) To minimize confusion we will use $N$ to refer to the parameter $n$ in the theorem being proved.

*Proof of Theorem 1.* In the argument below we choose $C_1$ large enough depending on $C_4$.

Given $N, m, r$, let $n_0 = \lceil n/2 \rceil$ and $n_1 = N - n_0$. Let $B$ be the largest integer such that $n_0 B \leq r$. Let $Y^0 = \{Bt : t \in [1, n_0]\}$. Consider the adversary for $G^n(m, r)$ that during the first $n_0$ steps loads $Y^0$ and then follows the optimal adversary strategy for the game $G^{n_1}(m|Y^0)$.

For the first part of Theorem 1, the hypothesis that $r \geq N2^{N-1}$ implies $B \geq 2^{n_1}$ so the first part of Lemma 3 applies. Note that $1 - \delta_0$ is at least $1/2$.

For the second part of Theorem 1, the hypothesis $r \geq C_1 m$ (and our freedom to choose $C_1$) implies $B \geq \lfloor C_1 m/n_0 \rfloor \geq C_1/\delta_0 - 1 \geq 12/\delta_0 + 1$ and so part (2) of Lemma 3 gives the desired lower bound. $\square$

*Proof of Theorem 2.* As mentioned above, we use $N$ to refer to the parameter $n$ in the theorem being proved. We describe an adversary strategy for the game $G^N(m,r)$. It will be convenient to assume that $m < (1+c)N$ where $0 < c \leq 1$ is the solution of $\ln((1+c)/c) = 28\ln(2/3)$ which implies $\ln(1/(1-\delta)) \geq 28\ln(2/3)$. This assumption is permitted since in the remaining case that $(1+c)N \leq m < 2n$, for $\delta = N/m$ we have that $\ln(1-\delta)$, $\delta$ and $\ln(1/\delta)$ can be bounded above and below by positive constants, and so Theorem 2 follows from Theorem 1.

The adversary works in phases. The adversary initially loads a set $Z_0$ of $N_0 = \lfloor m/3 \rfloor$ keys. This leaves $s_0 = m - N_0$ empty spaces in the array. In phase 1, the adversary loads $N_1 = \lfloor s_0/3 \rfloor$ additional items (according to a strategy described below), and sets $s_1 = s_0 - N_1$, which is the number of empty spaces in the array after phase 1. In general, after phase $j-1$ the number of empty spaces remaining will be $s_{j-1}$ and in phase $j$ the adversary loads $N_j = \lfloor s_{j-1}/3 \rfloor$ additional items. We will run the process for $p = \lfloor \ln(1-\delta)/7\ln(2/3) \rfloor - 1$ phases. By the choice of $c$, $p \geq 3$. For each $j \leq p$, we have $s_j \geq m(2/3)^{j+1} \geq m(1-\delta)^{1/7} \geq m(1-\delta) \geq m - N$, Hence, there are enough items to run the $p$ phases. We will show a lower bound for the cost of these phases. If not all items are loaded during the phases the additional items will only increase the cost.

Let $Z_{j-1}$ be the set of items loaded through the end of phase $j-1$ and $z_{j-1} = |Z_{j-1}|$. Thus $z_{j-1} + s_{j-1} = m$. Let $\delta_{j-1} = |Z_{j-1}|/m = \sum_{i=0}^{j-1} N_i/m$ be the density at the beginning of phase $j$. During phase $j$, the adversary uses the strategy for $G^{N_j}(m|Z_{j-1})$ provided by the adversary in Lemma 3. We need to verify that the conditions of the lemma are satisfied.

The parameter $n$ in the lemma is $N_j$ and we need this to be at least $C_4$. For each $j \in [1,p]$,

$$\begin{aligned} N_j &\geq m(2/3)^j/3 - 1 \geq m(1-\delta)^{1/7}/3 - 1 \\ &\geq m^{6/7}(m-N)^{1/7}/3 - 1 > N^{5/6}, \end{aligned} \qquad (2)$$

for $N$ large enough. This is at least $C_4$. The parameter $\delta_0$ in the lemma is $\delta_{j-1}$ and we need that this is in the interval $((\ln(N_j))^{-2}, 1 - N_j^{-1/5})$. For $N$ large enough $\delta_{j-1} \geq N_0/m \geq 1/4$ and the lower bound holds. For the upper bound on $\delta_{j-1}$, we have $\delta_{j-1} \leq (m-s_{j-1})/m \leq (m-3N_j))/m \leq 1 - 3N_j/2N \leq 1 - 3N_j/2N_j^{6/5} \leq 1 - N_j^{-1/5}$. Thus, $\delta_{j-1}$ satisfies the conditions for $\delta_0$ in Lemma 3.

We also need that in each phase $\text{mingap}(Z_{j-1}) \geq 1 + 12/\delta_{j-1}$. Since $\delta_{j-1} \geq 1/4$ it suffices that $\text{mingap}(Z_{j-1}) \geq 49$. By the remark following Lemma 3, we may assume that the value of mingap reduces by a factor of at most 49 in each phase, so in phase $j$ the value of mingap is at least $r/(N \cdot 49^{j-1})$ so it suffices that $r > N \cdot 49^p$. Since $49^p \leq (1/(1-\delta))^{C_2}$ for an appropriate constant $C_2$, the hypothesis on $r$ in the Theorem is sufficient.

We now show a lower bound on the cost of a single phase $j$. First, $N_j = \lfloor s_{j-1}/3 \rfloor = \lfloor m(1-\delta_{j-1})/3 \rfloor \geq m(1-\delta_{j-1})/4$, for $N$ large enough. We will use the fact that for any real $x \in [1/4, 1]$, $\ln(1/x) \leq 3(1-x)$. From Lemma 3 we obtain the following lower bound on the cost of the phase $j$:

$$\begin{aligned} N_j(\ln(N_j))^2 \frac{\delta_{j-1}^2(1-\delta_{j-1})}{C_4(\ln(1/\delta_{j-1}))^2} &\geq \frac{N_j}{1-\delta_{j-1}}(\ln(N_j))^2 \frac{1}{4^2 \cdot C_4} \frac{(1-\delta_{j-1})^2}{(\ln(1/\delta_{j-1}))^2} \\ &\geq \frac{m}{4}(\ln(N^{5/6}))^2 \cdot \frac{1}{4^2 \cdot C_4 \cdot 3^2} \\ &\geq N(\ln(N))^2/C_7, \end{aligned}$$

for some constant $C_7$.

Since the number of phases is more than $\ln(1/(1-\delta))/14\ln(3/2)$, we obtain the required lower bound. □

It remains to prove Lemma 3.

## 3 Some preliminaries for the lower bound

### 3.1 Segments, time intervals, key intervals, and lazy algorithms

We use the following terminology

- A *segment* is a subinterval of the set of cells $[1, m]$.

- A *time interval* is a subinterval of $[0, n]$.

- A *key interval* is a subinterval of the set $[1, r]$ of keys. If $Y \subseteq [1, r]$ is any set of keys, a $Y$-*interval* is a set of the form $Y \cap I$ where $I$ is a key interval.

Recall that at step $t$, $Rel^t$ denotes the set of keys relocated at step $t$. For $y \in Rel^t$ the *trail of $y$ at step $t$* is the segment $Trail^t(y)$ between $f^{t-1}(y)$ and $f^t(y)$; for $y^t$ it is just the location $f^t(y^t)$. The *busy region* at step $t$, denoted $B^t$ is the union over $y \in Rel^t$ of $Trail^t(y)$.

We say that an algorithm is *lazy* if $B^t$ is a segment. The following proposition says that we may restrict attention to lazy algorithms.

**Proposition 4** *Given any algorithm $A$ there is a lazy algorithm $A'$ such that for any initial key set $Y^0$ and any key sequence $y = (y^1, \ldots, y^n)$ the cost of $A'$ on $Y^0, y$ is at most the cost of $A$ on $Y^0, y$.*

**Proof:** The idea is that if the busy region $B^t$ is a union of two or more disconnected segments, then any relocation outside of the segment that contains $f^t(y^t)$ can be deferred until later.

To make this precise, the algorithm $A'$ keeps track of the storage function $g^t$ that would be produced by the algorithm $A$. Let $f^t$ be the array actually produced by $A'$.

Initially, prior to step 1, $f^0 = g^0$. At each step $t$, $A'$ updates $g^{t-1}$ to $g^t$ based on algorithm $A$. It then produces $f^t$ as follows: For a segment $T$ let $K(T)$ be the keys stored in $S$ under $f^{t-1}$. Let $S$ be the smallest segment containing $g^t(y^t)$ (the location chosen by $A$ to store $y^t$) with the property that $K(S) \cup \{y^t\}$ is the same as the set of elements stored in $S$ in $g^t$. The algorithm then defines $f^t$ so that every key in $K(S)$ is stored according to $g^t$, and every other key is stored according to $f^{t-1}$.

The definition of $S$ ensures that the busy region $B^t$ of $A'$ will be exactly the segment $S$. (Clearly $B^t \subseteq S$; to see that it is equal suppose there is a location $j$ in $S$ that is not in $B^t$. Without loss of generality $j$ is left of $f^t(y^t)$. Then if we shrink $S$ by moving its left endpoint to $j+1$ then the resulting segment contradicts the choice of $S$.)

Finally, we need to show that the cost of relocations by $A'$ is no more than the cost of relocations by $A$. For this, consider all relocations of a fixed key $y$. An easy induction shows that up through the end of any step $t$ the number of times $y$ was relocated by $A'$ is less than or equal to the number of times $y$ was relocated by $A$, with a strict inequality if $f^t(y) \neq g^t(y)$. □

Henceforth we assume that the algorithm is lazy, and refer to $B^t$ as the *busy segment at step $t$*.

# 4  Suitable gaps and segment table strategies

In this section we describe the high-level structure of the adversary strategy, and state several parametrized properties that we will use about the strategy.

During each step $t$ the adversary must choose a key $y^t$ to load into the array. For a set $Y$ of keys, a $Y$-*gap* is a pair $y_L < y_R$ of keys belonging to $Y$ such that no key of $Y$ has value in the key interval $(y_L, y_R)$. The *gap length* is $y_R - y_L$. Provided that the gap length is at least 2, there is always a key between $y_L$ and $y_R$ that is available to be loaded. We call such a gap *suitable*. A *suitable segment* is one that contains a suitable gap. Our adversary will choose a suitable segment, identify the largest suitable gap $(y_L, y_R)$ stored in the segment and select the key $\lfloor (y_L + y_R)/2 \rfloor$, which is the midpoint of the gap rounded down to the nearest integer. The segment (resp., gap) chosen by the adversary at step $t$ is referred to as the *chosen segment (resp., gap) at step $t$*.

To head off possible confusion, we emphasize that a gap refers to the set of possible key values between $y_L$ and $y_R$ and not to the region of the array in which the keys are stored.

The reader should think of step $t$ as consisting of the following sequence of events.

1. The configuration $(Y^{t-1}, f^{t-1})$ was specified prior to time step $t$. We refer to the associated configuration and density functions as *the configuration at the end of step $t-1$* or *at the beginning of step $t$*. We emphasize that the configuration at the beginning of step $t$ is $(Y^{t-1}, f^{t-1})$ and not $(Y^t, f^t)$.

2. During the first part of step $t$, the adversary selects a suitable segment $S$ (with respect to configuration $(Y^{t-1}, f^{t-1})$). We say that such a segment is *suitable for step $t$*.

3. The adversary chooses the largest gap in $S$ and lets $y^t$ be the rounded midpoint. $Y^t$ is set to be $Y^{t-1} \cup \{y^t\}$.

4. The algorithm selects the storage function $f^t$ for $Y^t$.

The choice of the suitable segment at step $t$ will depend on the configuration $(Y^{t-1}, f^{t-1})$. Intuitively, the adversary will select a suitable segment that is currently located in an area of the array that is relatively "crowded". For this purpose we fix a real parameter $\lambda > 0$ called the *weight parameter*, and define the *weight of a key* $y$ to be 1 if $y \in Y^0$ and $\lambda$ otherwise. Given a configuration $(Y, f)$, we define the following functions on segments $S \subseteq [1, m]$:

- The weight $w(S) = w(S, f)$ is the weight of all keys stored in $S$ under $f$.

- The density $\rho(S) = \rho(S, f)$ is $w(S)/|S|$. The density function provides a natural measure of crowding of $S$.

We write $w^{t-1}(S)$ and $\rho^{t-1}(S)$ for the weight and density of segment $S$ with respect to the configuration $(Y^{t-1}, f^{t-1})$.

We now describe how the weight parameter is chosen. The weight parameter depends on two things: $\operatorname{mingap}(Y^0)$, and an auxiliary parameter $\delta^*$, called the *density lower bound parameter*. (This will turn out to be a lower bound on the density of certain segments that arise in the definition of the adversary).

*Weight parameter specification.* Let $\delta^* > 0$ be the auxiliary density lower bound parameter (to be specified later). Let $Y^0$ be a set of keys with $\operatorname{mingap}(Y^0) \geq 1 + 4/\delta^*$.

$$\lambda = \begin{cases} 1 & \text{if } \operatorname{mingap}(Y^0) \geq 2^n, \\ \delta^*/2 & \text{if } \operatorname{mingap}(Y^0) < 2^n. \end{cases} \tag{3}$$

This choice of parameters gives the following lemma, which establishes sufficient conditions on a segment to contain a suitable gap.

**Lemma 5** *Let $\delta^*$ and $\lambda$ be as given in the weight parameter specification. Let $\mu_0 = \text{mingap}(Y^0)$ and assume $\mu_0 \geq 1 + 4/\delta^*$. Let $t \in [1,n]$ be any step and let $S$ be any segment whose weight (with respect to $w^{t-1}$) is at least 2 and whose density is at least $\delta^*$. Then $S$ is suitable for step $t$.*

**Proof:** First consider the case that $\mu_0 \geq 2^n$, and so $\lambda = 1$. An easy induction shows that the minimum gap in $Y^t$ is at least $2^{n-t}$, and so after step $t-1$ the minimum gap is at least 2, so every gap is suitable. Since $S$ has weight at least 2 it contains at least one gap.

Next consider the case that $\mu_0 < 2^n$. Let $A$ be the set of keys from $Y^0$ stored in $S$ after step $t-1$ and $B$ be the set of other keys stored in $S$. Let $a = |A|$ and $b = |B|$. We first claim that $a \geq 2$. The weight of $S$ is $a + b\lambda = |S|\rho(S) \geq (a+b)\delta^*$ which implies $a > b(\delta^* - \lambda) = b\lambda$, which implies $a > (a+b\lambda)/2 = w^{t-1}(S)/2$ which is at least 1 by hypothesis. Thus $a \geq 2$.

Let $\min_A$ and $\max_A$ be the smallest and largest keys in $A$. Suppose for contradiction that there is no suitable gap between $\min_A$ and $\max_A$. Then all of the $\max_A - \min_A + 1$ keys in the range $[\min_A, \max_A]$ must have been loaded already. There are $a-1$ gaps between keys of $A$, each of size at least $\mu_0$ so we must have $b \geq (a-1)(\mu_0 - 1) \geq (a/2)4/\delta^* = a/\lambda$, which contradicts $a > b\lambda$. □

The adversary we describe will identify a segment satisfying the conditions of Lemma 5 (and other conditions as well.) The strategy is based on a structure called a *segment table*. A segment table is an array with $n$ columns (one for each step) whose entries are array segments. The entries of the table are colored *green* or *red*. The rows of the segment table are called *levels* and the number of levels, which we normally denote by $d$, is the *depth*. The level index increases from the top to the bottom of the array. A level-step pair $(i,t) \in [1,d] \times [1,n]$ is called a *site*, and the table entry (segment) at site $(i,t)$ is denoted $S_i^t$. It is sometimes convenient to consider $(0,t)$ for $t \in [1,n]$ to be a site even though there is no corresponding table entry. The segment table must satisfy:

- The segments down each column are nested: $S_1^t \supset \cdots \supset S_d^t$.

- The color of an entry in column $i$ is defined as follows: $S_i^t$ is green if $B^t \subseteq S_i^t$ (recall that $B^t$ is the busy segment at step $t$) and is red otherwise. Together with the nesting property this implies that each column consists of a (possibly empty) sequence of green entries followed by a (possibly empty) sequence of red entries.

- If segment $S_i^t$ is green then $S_i^{t+1} = S_i^t$.

For each level $i$, we partition $[1,n]$ into intervals called *i-epochs*, where each red site at level $i$ marks the end of an $i$-epoch. The last $i$-epoch (which ends at step $n$) is called the *terminal i-epoch* and the others are *non-terminal*.

An $i$-epoch $E$ is identified with the set $\{i\} \times E$ of sites. All sites but the last site in the epoch are green. For a non-terminal epoch the last site is red, for a terminal epoch the last site may be red or green.

For an $i$-epoch $E$, the left endpoint of $E$ is called the *starting time* and is denoted $s(E)$ and the right endpoint is called the *closing time* and is denoted $c(E)$.

By the properties of the segment table, every site of $E$ is associated to the same segment, which is denoted $S_i^E$. Since for each column every site above a green site is green, epoch $E$ is a subset of an epoch at level $i-1$, so all sites in $\{i-1\} \times E$ are associated to the same segment, denoted $S_{i-1}^E$.

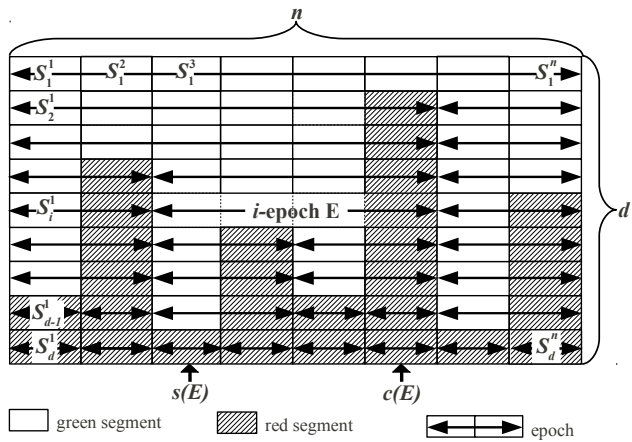

Figure 1: *Segment table.*

For a site $(i,t)$ we write $E(i,t)$ for the $i$-epoch containing $t$ and $s(i,t)$ and $c(i,t)$ for the starting time and closing time of that epoch.

We will specify an adversary that builds the segment table by constructing column $t$ during step $t$. For our adversary, the events of step $t$ described earlier can be refined as follows:

- The configuration $(Y^{t-1}, f^{t-1})$ was specified prior to time step $t$.

- During the beginning portion of time step $t$, the adversary chooses the segments for column $t$. This selection is based on the configuration $(Y^{t-1}, f^{t-1})$ and functions derived from it such as $\rho^{t-1}$. The segments will be chosen in such a way that $S_d^t$ satisfies the hypothesis of Lemma 5 (with respect to $(Y^{t-1}, f^{t-1})$), and therefore contains a suitable gap.

- The adversary chooses the largest gap in $S_d^t$ and lets $y^t$ be the approximate midpoint. $Y^t$ is set to be $Y^{t-1} \cup \{y^t\}$.

- The algorithm selects the storage function $f^t$ for $Y^t$.

- The choice of $f^t$ together with the previous storage function $f^{t-1}$ determines the busy segment $B^t$.

- Each segment $S_i^t$ in column $t$ is colored green (if $S_i^t$ contains $B^t$) or red (if $S_i^t$ does not contain $B^t$).

A procedure for the adversary to choose column $t$ given $(Y^{t-1}, f^{t-1})$ completely determines an adversary strategy. We call such a strategy a *segment table* strategy.

We will specify and analyze a particular segment table strategy. We begin by identifying some additional properties we want our table to satisfy, and then show how these properties lead to a lower bound on the cost incurred by the algorithm.

We use the segment table to help account for the cost of the relocations done by the algorithm. We partition $Rel^t$ (the set of keys relocated at step $t$) into subsets $Q_0^t, \ldots, Q_d^t$ as follows: For $i \geq 1$, $Q_i^t$ is the set of $y \in Rel^t$ such that $f^{t-1}(y) \in S_i^t - S_{i+1}^t$ (i.e. $S_i^t$ is the smallest segment in column $t$ that contains the location that $y$ was moved from). We include $y^t$ in $Q_d^t$. $Q_0^t$ is the set of those $y$ that were moved from a location outside of $S_1^t$. Let $q_i^t = |Q_i^t|$. For an $i$-epoch $E$ at level $i$ we define $q_i^E = \sum_{t \in E} q_i^t$, which is the total cost of relocations associated to $E$. Thus the cost incurred by the algorithm is $\sum_{i=0}^{d} \sum_E q_i^E$, where the inner sum is over all $i$-epochs.

We are now ready to state the desired properties of the segment table. These properties depend on three *strategy parameters*: $\delta^*$ (the density lower bound parameter introduced earlier, which determines the weight parameter $\lambda$), $\gamma$ and $\alpha$. These parameters will be chosen later.

(P1) The number of levels $d$ is an integer greater than or equal to 8.

(P2) All segments have size at most $n/2$.

(P3) For each $t$ and $i \geq 2$, $|S_i^t| \leq |S_{i-1}^t|/2$. (Segment sizes decrease by at least a constant factor down columns)

(P4) All segments have size at least $1/\gamma$.

(P5) $\rho^{t-1}(S_1^t) \geq \delta_0 e^{-\alpha}$ and for $i \geq 2$, $\rho^{t-1}(S_i^t) \geq e^{-\alpha}\rho^{t-1}(S_{i-1}^t)$. (Segment densities do not decrease much down columns).

(P6) Every segment in the table has density at least $\delta^*$

(P7) For any non-terminal $i$-epoch $E$ with starting time $s$, $q_i^E \geq \frac{1}{8}w^{s-1}(S^E)$, that is, the relocation cost associated with epoch $E$ is at least a $1/8$-fraction of the weight of the associated segment $S^E$ at the start of the epoch.

One of the constraints we will impose on the parameters (condition (I1) below) is $\delta^* \geq 2\gamma$. With this constraint, (P4) and (P6) together with Lemma 5 guarantee that the segments $S_d^t$ are suitable.

We will prove two lemmas. The first lemma (Lemma 6) gives a lower bound on the cost incurred by the algorithm against a segment table strategy that satisfies the above properties, in terms of the parameters $\gamma, \alpha$ and $\delta^*$ in the properties. The second (Lemma 11) shows that there is a segment table strategy that satisfies the above properties with suitable values of the parameters. Finally, in section 6.4 we use these two lemmas to prove Lemma 3.

## 5 A segment table strategy gives a good lower bound

In this section we prove a lemma that proves a lower bound on the cost incurred by an algorithm based on a segment table strategy. The lemma encapsulates and extends the main accounting argument of Dietz et al. [DSZ05a, Zha93], which they used to prove an $\Omega(\ln^2(n))$ amortized lower bound for the special case of *smooth* algorithms.

**Lemma 6** *Let $m, n, n_0, \delta_0$ and $Y^0$ be as in Lemma 3. Let $\alpha, \delta^*, \gamma$ be positive parameters and let $\lambda$ be the associated weight parameter as defined earlier. If a segment table strategy produces a segment table with $d$ levels satisfying (P1)-(P7) then the cost incurred by the algorithm satisfies*

$$\chi^n(m|Y^0) \geq \frac{\delta^* \lambda n d^2}{500(\alpha d + \gamma\lambda + 1 - \delta_0)}. \tag{4}$$

**Remark.** When we choose values for these parameters, in the case that $\delta_0$ is a constant in $(0, 1)$, $\delta^*$ and $\lambda$ will be bounded below by positive constants, and the denominator will be bounded above by a positive constant. Also $d$ will be $\Theta(\ln(n))$, yielding an $\Omega(n \ln(n)^2)$ bound.

**Proof:** Let $\chi$ denote the cost of a given algorithm against a segment table-based strategy as in the lemma. As noted above $\chi \geq \sum_{i=0}^{d} \sum_E q_i^E$. Let $E^+$ denote the set consisting of the largest $\lceil |E|/2 \rceil$ time steps belonging to $E$. We will spread the cost of epoch $E$ among the sites corresponding to $E^+$. It is easy to check that for each $t \in E^+$, $|E^+| \leq (t - s(E) + 1)$ and thus:

$$q_i^E \geq \sum_{t \in E^+} q_i^E / (t - s(E) + 1).$$

Say that a site $(i, t)$ is *chargeable* if (i) $i \in [2, d-1]$, (ii) the $i$-epoch $E(i, t)$ containing $t$ is a non-terminal epoch (not the final epoch at level $i$), and (iii) $t \in E(i, t)^+$. Let $CS$ denote the set of chargeable sites. From (P7) and (P6) we have that for an $i$-epoch $E$, $q_i^E \geq \frac{1}{8} w^{s(E)-1}(S^E) \geq \delta^* |S^E|/8$, and so:

$$\chi \geq \frac{\delta^*}{8} \sum_{(i,t) \in CS} \frac{1}{(t - s(i,t) + 1)/|S_i^t|},$$

where $s(i, t)$ is the starting time of the $i$-epoch containing $t$.

We use following standard fact (the arithmetic-harmonic mean inequality):

**Proposition 7** *For $a_1, a_2, \ldots, a_p, k \geq 0$.*

$$\sum_{i=1}^{p} \frac{1}{a_i} \geq \frac{p^2}{\sum_{i=1}^{p} a_i}.$$

The length of every epoch $E$ is at most $|S^E|$ (since at most $|S^E|$ keys can be stored in $S^E$ before some keys are moved outside of $S^E$) and so (P2) implies that all epochs (including terminal epochs) are length at most $n/2$. Thus at each level the union of the non-terminal epochs has size at least $n/2$ and thus the number of chargeable sites at a given level $i$ is at least $n/4$. Thus $|CS| \geq n(d-2)/4$. Applying the proposition together with (P1) gives

$$\chi \geq \frac{\delta^* n^2 (d-2)^2}{128 \sum_{(i,t) \in CS} (t - s(i,t) + 1)/|S_i^t|} \geq \frac{\delta^* n^2 d^2}{250 \sum_{(i,t) \in CS} (t - s(i,t) + 1)/|S_i^t|} \quad (5)$$

It remains to upper bound the sum in the denominator. Since for every $(i, t)$ the term in the denominator is nonnegative, it suffices to bound the extended sum where $i \in [2, d-1]$ and $t \in [1, n]$. For each fixed $t \in [1, n]$, we can bound the terms of this sum corresponding to $t$ by:

$$\sum_{i=2}^{d-1} 1/|S_i^t| + \sum_{i=2}^{d-1} \frac{t - s(i,t)}{|S_i^t|}. \quad (6)$$

We bound the first sum using (P3) and (P4):

$$\sum_{i=2}^{d-1} \frac{1}{|S_i^t|} \leq \frac{1}{|S_d^t|} \sum_{j \geq 0} (1/2)^j \leq 2\gamma.$$

To bound the second sum we claim that for any $i$-epoch $H$ containing $t$ and any time $u$ satisfying $s(H) \leq u \leq t$:

$$\frac{t-u}{|S^H|} = \frac{1}{\lambda}(\rho^{t-1}(S^H) - \rho^{u-1}(S^H)). \tag{7}$$

This follows since all sites in $\{i\} \times (H - \{c(H)\})$ are green, so the set of keys stored in $S^H$ after step $t-1$ consists of those stored after step $u-1$ together with those loaded during steps $u, \ldots, t-1$ and therefore the density increases by exactly $\lambda(t-u)/|S^H|$.

Fix $i \in [2, d-1]$. Let $E = E(i,t)$ and $F = E(i-1,t)$. Applying (7) with $H = E$ and $u = s(i,t)$ gives an upper bound on the summand of the second sum in (6) as a sum of differences. It is hard to analyze the sum of these differences directly, so we do a little more manipulation. Since $E \subseteq F$, we can apply (P3) together with (7) with $H = F$ and $u = s(i,t)$ to get:

$$\frac{t-s(i,t)}{|S_i^t|} \geq 2\frac{t-s(i,t)}{|S^F|} = \frac{2}{\lambda}(\rho^{t-1}(S^F) - \rho^{s(i,t)-1}(S^F)). \tag{8}$$

Subtracting (8) from twice (7) (with $H = E$ and $u = s(i,t)$), and rearranging terms gives:

$$\frac{t-s(i,t)}{|S_i^t|} \leq \frac{2}{\lambda}[(\rho^{t-1}(S^E) - \rho^{t-1}(S^F)) + (\rho^{s(i,t)-1}(S^F) - \rho^{s(i,t)-1}(S^E))]. \tag{9}$$

When we sum this over $i = 2$ to $d-1$ the first part of the sum telescopes to $\frac{2}{\lambda}(\rho^{t-1}(S^{E(d-1,t)}) - \rho^{t-1}(S^{E(1,t)}))$. By (P5), this is at most $\frac{2}{\lambda}(1 - e^{-\alpha}\delta_0) \leq \frac{2}{\lambda}(1 - \delta_0 + \alpha)$.

For the second part we have by (P5) (since $E$ starts at time $s(i,t)$):

$$\frac{2}{\lambda}(\rho^{s(i,t)-1}(S_{i-1}^t) - \rho^{s(i,t)-1}(S_i^t)) \leq \frac{2}{\lambda}(1 - e^{-\alpha})\rho^{s(i,t)-1}(S_{i-1}^{s(i,t)}) \leq \frac{2\alpha}{\lambda}.$$

Summing over $i \in [2, d-1]$ gives at most $\frac{2}{\lambda}\alpha(d-2)$.

Combining the three parts of the sum of the denominators we get an upper bound of $\frac{2n}{\lambda}(\alpha(d-1) + \gamma\lambda + 1 - \delta_0) \leq \frac{2n}{\lambda}(\alpha d + \gamma\lambda + 1 - \delta_0)$. This yields the desired lower bound $\chi \geq \delta^*\lambda n d^2/500(\alpha d + \gamma\lambda + 1 - \delta_0)$. □

## 6 Construction of a good segment table adversary

In this section we give a construction for a segment table adversary. The construction takes two parameters: the number of levels $d$ and an auxiliary *potential function parameter* $\kappa > 0$. In Lemma 11 we prove that for a particular choice of parameters $d$ and $\kappa$ the adversary is well-defined and satisfies the properties (P1)-(P7) for specific choices of $\alpha, \gamma, \delta^*$.

To specify the strategy we need a rule which given a step $t$ and the first $t-1$ columns of the table (including the colors) selects the segments for column $t$. To define the rule, it will be convenient to augment the segment-table by defining an auxiliary segment $T_i^t$ at each site $(i,t)$, which we view as sharing the site $(i,t)$ with $S_i^t$ in the segment-table. The segments $T_i^t$ will satisfy:

$$T_1^t \supset S_1^t \supset T_2^t \supset S_2^t \supset \cdots \supset T_d^t \supset S_d^t$$

Our adversary will select the $T_i^t$ and $S_i^t$ in the order above. To describe how this is done, we need some additional definitions and observations.

- The segment $W^t$ is defined as follows: Divide $[1, m]$ into segments from left to right $A_1, A_2, \ldots, A_r$ where each $A_i$ has size $\lceil n/2 \rceil$ except the last $A_r$, whose size is between $\lceil n/2 \rceil$ and $2\lceil n/2 \rceil \leq n+1$. Take $W^t$ to be that segment $A_i$ that maximizes $\rho^{t-1}(A_i)$ (breaking ties arbitrarily). Observe that $\rho^{t-1}(W^t) \geq \rho^{t-1}([1, m]) \geq \delta_0$.

- For a segment $T$, **middle**$(T)$ is defined as follows: Break $T$ into three segments from left to right, $L, M, R$ where $|L| = |R| = \lfloor |T|/3 \rfloor$. **middle**$(T)$ is the segment $M$.

Let $S$ be a segment and $\rho^{t-1}$ be the density function at the end of step $t-1$. Let $\kappa > 0$.

- $S$ is $\kappa$-*upper balanced* (with respect to $\rho^{t-1}$) if every subsegment of size at least $|S|/4$ has density at most $\rho^{t-1}(S)4^\kappa$.

- $S$ is $\kappa$-*lower balanced* if every subsegment of size at least $|S|/4$ has density at least $\rho^{t-1}(S)(1/4)^\kappa$.

- Define the $\kappa$-*potential of segment* $S$ to be $\phi_\kappa^{t-1}(S) = |S|\rho^{t-1}(S)^{1/\kappa}$.

- For a segment $T$, **densify**$_\kappa^{t-1}(T)$ is the subsegment $D$ that maximizes $\phi_\kappa^{t-1}(D)$ (breaking ties arbitrarily).

We note the following easy facts.

**Proposition 8** *Let $T$ be an arbitrary segment and $t \in [1, n]$.*

1. *The size of $|T|$ is at least its potential (since $\rho^{t-1}(T) \leq 1$).*

2. $\rho^{t-1}(\mathbf{densify}_\kappa^{t-1}(T)) \geq \rho^{t-1}(T)$.

3. *If $T$ is not $\kappa$-upper balanced (with respect to $\rho^{t-1}$) and $D$ is a subsegment of $T$ that violates the conditions of $\kappa$-upper balance then $\phi_\kappa^{t-1}(D) > \phi_\kappa^{t-1}(T)$. Thus, since $\mathbf{densify}_\kappa^{t-1}(T)$ has no subsegment with larger $\phi_\kappa^{t-1}$, it must be $\kappa$-upper balanced.*

Finally, we define **balance**$_\kappa^{t-1}(T)$ to be the subsegment **middle**(**densify**$_\kappa^{t-1}(T)$) of $T$. The properties of **balance**$_\kappa^t(T)$ that we need are given by the following lemma.

**Lemma 9** *Assume $\kappa \leq 1/24\ln(4)$. Fix a step $t \in [1, n]$. Let $T$ be a segment, let $D = \mathbf{densify}_\kappa^{t-1}(T)$ and $S = \mathbf{balance}_\kappa^{t-1}(T) = \mathbf{middle}(D)$. Assume $|S| \geq 4$. Then*

- $\rho^{t-1}(S) \geq e^{-24\ln(4)\kappa}\rho^{t-1}(D) \geq e^{-24\ln(4)\kappa}\rho^{t-1}(T)$.

- $S$ *is $25\kappa$-lower balanced with respect to $\rho^{t-1}$.*

- $\phi_\kappa^{t-1}(S) \geq \phi_\kappa^{t-1}(T)e^{-(24\ln(4)+\ln(3))}$.

We give the proof in section 6.1

We are now ready to describe the adversary strategy for selecting segments in column $t$ at step $t$. For $t \geq 2$, this selection will depend on the segments and coloring of the previous column. Recall that the red-green coloring of the previous column $t-1$ is determined by the action of the algorithm in response to $y^{t-1}$, and that $B^{t-1}$ is the busy segment at step $t-1$, which is the minimal segment of the array in which all rearrangements occurred. The adversary depends on the parameter $d$ (the number of levels) and the potential function parameter $\kappa > 0$.

**Adversary**$(d, \kappa)$

- Preservation rule: If $t \geq 2$ then for $i = 1, \ldots, d$, if $(i, t-1)$ is green then $T_i^t = T_i^{t-1}$ and $S_i^t = S_i^{t-1}$. (Copy the corresponding segments from the previous column.)

- Let $j^t$ be the first level $i$ to which the preservation rule does not apply. This is the *t-critical level*.

- (Rebuilding Rule)

  For $i = j^t, \ldots, d$:

  - Determine $T_i^t$
    * If $i$ is the critical level then:
      · If $i = 1$ then $T_i^t = W^t$.
      · If $i > 1$ then $T_i^t = T_i^{t-1} \cup B^{t-1}$.
    * If $i$ is not the critical level then $T_i^t = \mathbf{middle}(S_{i-1}^t)$.
  - Determine $S_i^t$: $S_i^t = \mathbf{balance}^{t-1}(T_i^t)$.

**Remark.** In the case that $i$ is the critical level and $i > 1$, $T_i^t$ is defined to be the union of two segments. This union is required to be a segment, and for this we need $T_i^{t-1} \cap B^{t-1} \neq \emptyset$. But this is clear since $B^{t-1}$ includes the locations where the selected gap $y_L^{t-1}, y_R^{t-1}$ was stored, and those locations are in $T_i^{t-1}$.

In section 6.1 we prove Lemma 9. In section 6.3 we state and prove a lemma that shows that Adversary$(d, \kappa)$ satisfies the desired properties (P1)-(P7) for particular choices of $\alpha, \gamma, \delta^*$.

## 6.1 Proof of Lemma 9

We now turn to verifying the needed properties of the function $\mathbf{balance}_\kappa^t$.

Throughout this section, we fix $t$. We omit the superscript $t-1$ from **balance**, **densify**, $\rho, \phi$ and $w$. Also, we fix $\kappa$ and omit the subscript $\kappa$ from **balance** and **densify**.

As noted earlier, $D = \mathbf{densify}(T)$ is $\kappa$-upper balanced.

**Claim 10** *Let $U$ be a subsegment of $S$ of size at least $|S|/4$. Then $\rho(U) \geq (1/4)^{24\kappa} \rho(D)$.*

Given the claim, we deduce the lemma. For the first part, we apply the claim with $S = U$ and note that $\rho(D) \geq \rho(T)$. For the second part we combine the claim with the fact that $\rho(D) \geq \rho(S)(1/4)^\kappa$ (which holds since $|S| \geq |D|/3$ and $D$ is $\kappa$-upper balanced.) For the third part of the lemma, we note that $\phi(S)/\phi(T) \geq \phi(S)/\phi(D) \geq \frac{1}{3}(\rho(S)/\rho(D))^{1/\kappa}$, and apply the first inequality of the first part.

So it suffices to prove the claim. The set $D - U$ consists of 2 segments $L$ (on the left) and $R$ (on the right). The weight of $D$ can be written as:

$$|D|\rho(D) = |L|\rho(L) + |R|\rho(R) + |U|\rho(U),$$

which implies:

$$\rho(U) = \frac{1}{|U|}(|D|\rho(D) - |L|\rho(L) - |R|\rho(R)).$$

Since $|S| \geq 4$ and $S = \mathbf{middle}(D)$ it follows that $|D| \geq 8$ and $|S| \leq |D|/2$. Since $U \subseteq S = \mathbf{middle}(D)$, we have $|L|, |R| \geq |D|/4$ and since $D$ is $\kappa$-upper balanced, $\rho(L), \rho(R) \leq 4^\kappa \rho(D)$. So:

$$\begin{aligned}
\rho(U) &\geq \frac{\rho(D)}{|U|}(|D| - (|L| + |R|)4^{\kappa}) \\
&\geq \frac{\rho(D)}{|U|}(|D|4^{-\kappa} - |L| - |R|) \\
&= \frac{\rho(D)}{|U|}(|D|4^{-\kappa} - |D| + |U|) \\
&= \frac{\rho(D)}{|U|}(|U| - |D|(1 - e^{-\ln(4)\kappa})) \\
&\geq \rho(D)(1 - \frac{|D|}{|U|}\ln(4)\kappa) \\
&\geq \rho(D)(1 - 12\ln(4)\kappa) \geq \rho(D)4^{-24\kappa},
\end{aligned}$$

where the final inequality uses the hypothesis that $\kappa \leq 1/(24\ln(4))$ and the inequality $(1-x) \geq e^{-2x}$ for $x \leq 1/2$.

## 6.2 Setting the parameters

The adversary strategy depends on parameter $\kappa$ and $d$. The properties we need involve parameters $\alpha, \gamma, \delta^*$.

We will need the following constants

- $C_5$, needed in Lemma 12.
- $C_0$, which is a lower bound on $n$ (chosen after $C_5$ is chosen) imposed to make various conditions hold.

We impose the following hypotheses on $n$ and $\delta_0$.

- (A1) $n \geq C_0$
- (A2) $\delta_0 \in (\ln(n)^{-2}, 1 - n^{-1/5})$.

We set the parameters as follows:

$$\kappa = \frac{2\ln(1/\delta_0)}{\ln(n)} \tag{10}$$

$$d = \lfloor \frac{1-\delta_0}{8C_5 \ln(1/\delta_0)} \ln(n) \rfloor \tag{11}$$

$$\alpha = 2C_5\kappa = \frac{4C_5 \ln(1/\delta_0)}{\ln(n)} \tag{12}$$

$$\gamma = n^{-1/4} \tag{13}$$

$$\delta^* = \delta_0 e^{\delta_0 - 1}. \tag{14}$$

**Lemma 11** *Let $m, n, n_0, Y^0, \delta_0, \mu_0$ be as in Lemma 3. Let the parameters be set according to (10)-(14). Let $\lambda$ be the weight parameter as previously specified. Assume (A1) and (A2). Then* **Adversary**$(d, \kappa)$ *satisfies (P1)-(P7).*

As one would expect, the choice of parameters is dictated by two considerations. During the proof of Lemma 11 various inequalities involving the parameters will be needed. The parameters must satisfy these. Subject to these inequalities, we seek to (approximately) maximize the expression in the lower bound of Lemma 6.

To isolate some of the technical computations, before presenting the proof of Lemma 11, we collect together the inequalities involving the parameters that are needed in the proof. We then explain how the parameters were chosen to satisfy these inequalities and (approximately) maximize the lower bound of Lemma 6.

(I1) $\delta^*/\gamma \geq 2$. (This will ensure that $S_d^t$ satisfies the hypotheses of Lemma 5 and thus has suitable gap.)

(I2) $\kappa \geq 2\ln(1/\delta_0)/\ln(n)$. This is needed to prove the lower bound (16) on the $\phi^{t-1}$-value of the set $T_1^t$.

(I3) $d \leq \frac{1}{2C_5}\ln(\gamma\sqrt{n})$. Together with (I2) this is used to prove (25) and (23) which give lower bounds on the $\phi^{t-1}$-value of all segments in column $t$, which directly implies (P4).

(I4) $\gamma \leq 1/4$. With (P4) this implies that every segment $S_i^t$ has size at least 4. This is a hypothesis of Lemma 12, where it is used in order to apply Lemma 9.

(I5) $d\kappa \leq \frac{1}{2C_5}\ln(\delta_0/\delta^*)$. This inequality is used to prove (P6), via (24).

(I6) $\kappa \leq \frac{1}{24\ln(4)}$ (for Lemma 9) and $\kappa \leq 1/50$ (for the third claim in the proof of (P7)).

(I7) $\alpha \geq 2C_5\kappa$. This is used to prove (P5) via (18) and (19).

The choice of parameters was made based on these inequalities as follows:

1. We will choose parameters so that the denominator of (4) is $O(1-\delta_0)$. For this we will need that $\alpha d = O(1-\delta_0)$ and $\gamma\lambda = O(1-\delta_0)$.

2. The parameter $\gamma$ is involved in inequalities (I1),(I3), and (I4). These inequalities leave a lot of room, and we choose $\gamma = n^{-1/4}$. Then (I4) holds, (I1) becomes $\delta^* \geq 2n^{-1/4}$, (I3) becomes $d \leq \ln(n)/8C_5$. Also, by the restriction (A2) on $\delta_0$ we have $\gamma = O(1-\delta_0)$ (as needed for the denominator of (4)).

3. We want $d$ to be large so we want $\alpha$ to be small. So we set (I7) to equality to determine $\alpha$ (as a function of $\kappa$).

4. We choose $\kappa$ as small as possible by making (I2) an equality. Since $n$ is sufficiently large, (I6) holds.

5. Having chosen $\alpha$ and $\kappa$, to have $\alpha d = O(1-\delta_0)$, we want $d = O(1-\delta_0)/\alpha$. We also need (I3). We can satisfy both by taking $d = \lfloor \frac{1-\delta_0}{8C_5\ln(1/\delta_0)}\ln(n) \rfloor$, noting that $1-\delta_0 \leq \ln(1/\delta_0)$ for all $\delta_0 \in (0,1)$.

6. In order to make (I5) hold we need $\delta^*$ small enough but for the lower bound we want $\delta^*$ large. So we choose $\delta^*$ as large as possible subject to (I5). Note that for this choice (I1) holds since $1/\gamma \geq e/\delta_0$ which is at most $e\ln(n)^2$ by (A2).

We now proceed to the proof of Lemma 11. In the proof we refer only to the inequalities (I1)-(I7) and not to the actual values of the parameters,

## 6.3 The adversary satisfies the required properties

In this section we prove that $\mathbf{Adversary}(d, \kappa)$ satisfies (P1)-(P7) for a suitable choice of parameters.

First we prove a lemma that relates the $\rho$ and $\phi$ values of the segments in the segment table.

**Lemma 12** *There are positive constants $C_5$ and $C_0$ such that the following holds for $\mathbf{Adversary}(d, \kappa)$ provided that $n \geq C_0$ and (I1)-(I7) hold. Suppose that for each $t \in [1, n]$ and $i \in [1, d]$ we have $|S_i^t| \geq 4$. For each $t \in [1, n]$*

$$\rho^{t-1}(T_1^t) \geq \delta_0 \tag{15}$$

$$\phi^{t-1}(T_1^t) \geq \frac{\sqrt{n}}{2}. \tag{16}$$

*and for all $i \geq 2$:*

$$\text{if } t \text{ starts an } i\text{-epoch then } S_i^t \text{ is } 25\kappa\text{-balanced with respect to } \rho^{t-1}. \tag{17}$$

$$\rho^{t-1}(S_i^t) \geq \rho^{t-1}(T_i^t)e^{-C_5\kappa} \tag{18}$$

$$\rho^{t-1}(T_{i+1}^t) \geq \rho^{t-1}(S_i^t)e^{-C_5\kappa} \quad \text{if } i \leq d-1 \tag{19}$$

$$\phi^{t-1}(S_i^t) \geq \phi^{t-1}(T_i^t)e^{-C_5} \tag{20}$$

$$\phi^{t-1}(T_{i+1}^t) \geq \phi^{t-1}(S_i^t)e^{-C_5} \quad \text{if } i \leq d-1 \tag{21}$$

**Proof:** We prove these statements by induction on $t$, and for fixed $t$ by induction on $i$.

We will repeatedly use the following easy fact:

**Proposition 13** *Let $S \subseteq S'$ be segments and $s < t$ be steps. Suppose that for all steps $r \in [s, t-1]$, the busy segment $B^r$ is a subset of $S$. Then $\rho^r(S)$, $\phi^r(S)$, $\rho^r(S)/\rho^r(S')$ and $\phi^r(S)/\phi^r(S')$ are all nondecreasing as a function of $r \in [s, t-1)$*

**Proof:** This follows from: At each step in $[s, t-1]$, both $w^r(S)$ and $w^r(S')$ increase by $\lambda$ and $w^r(S) \leq w^r(S')$. □

Proof of (15) and (16). Suppose first that $t$ is the starting time of the 1-epoch containing $t$. Then $T_1^t = W^t$, and $\rho^{t-1}(T_1^t) \geq \delta_0$. We have $\phi^{t-1}(T_1^t) \geq (n/2)\delta_0^{1/\kappa}$ and by (I2) this is at least $\sqrt{n}/2$. For $t$ not the starting time of the epoch we use Proposition 13.

For the proofs of the remaining parts, we will need to apply Lemma 9 with $T = T_i^t$ and $S = S_i^t$. The hypotheses of Lemma 9 follows from (I6) and the hypothesis $|S_i^t| \geq 4$ of the present lemma.

Proof of (17). This follows immediately from the second part of Lemma 9.

Proof of (18) and (20). In the case that $t$ starts an $i$-epoch these follow immediately from the first and third parts of Lemma 9 provided that we choose $C_5 \geq 24 \ln(4) + \ln(3)$. For a step that does not start an $i$-epoch, this follows from Proposition 13.

Proof of (19) and (21). Let $E$ be the $i$-epoch containing $t$. In the case that $t$ starts an $i$-epoch this follows from the second part of Lemma 9. If $t$ does not start an $i$-epoch, let $s$ be the starting time of the epoch. We cannot apply Proposition 13 directly because, while the segment $S_i^t = S_i^s$, it may not be true that $T_{i+1}^t = T_{i+1}^s$ because there may have been one or more new $i+1$-epochs started. However, at each of these new $i+1$-epochs, $i+1$ was the critical level (since the $i$-epoch did not end) which means that the set $T_{i+1}^t$ is equal to the union of $T_{i+1}^s$ and all of the busy segments $B^r$ for $s \leq r \leq t-1$. This is a subsegment of $S_i^t = S_i^s$ that contains $T_{i+1}^s$. Hence we can apply the second part of Lemma 9 to get that at the beginning of the epoch $\rho^{s-1}(T_{i+1}^t)$ is at least $e^{-C_5\kappa}\rho^{s-1}(S_i^s)$, provided that $C_5 \geq 25\ln(4)$. Now we can apply Proposition 13 to show that the same inequality holds for $\rho^{t-1}$ (keeping in mind that $S_i^t = S_i^s$). Since $|T_{i+1}^t|/|S_i^t| \geq 1/3$ we also get (21), provided that $C_5 \geq 25\ln(4) + \ln(3)$. □

*Proof of Lemma 11.*

Using Lemma 12 repeatedly we have by induction on $i = 1, \ldots, d$ for fixed $t \in [1, n]$, that:

$$\rho^{t-1}(T_i^t) \geq \delta_0 e^{(2-2i)C_5\kappa} \tag{22}$$

$$\phi^{t-1}(T_i^t) \geq \frac{\sqrt{n}}{2}e^{(2-2i)C_5} \geq 1/\gamma \tag{23}$$

$$\rho^{t-1}(S_i^t) \geq \delta_0 e^{(1-2i)C_5\kappa} \geq \delta^* \tag{24}$$

$$\phi^{t-1}(S_i^t) \geq \frac{\sqrt{n}}{2}e^{(1-2i)C_5} \geq 1/\gamma. \tag{25}$$

The final inequality of (24) follows from (I5). The final inequalities of (23) and (25) follow from (I3). Note that the final inequality of (25) and (I4) imply that as we proceed to level $i$ in the induction, the hypothesis $|S_i^t| \geq 4$ of Lemma 12 holds at each step.

*Proof of Property (P1).* By definition $d$ is an an integer; we need to verify that $d \geq 8$. Consider the definition of $d$ given in (11). If $\delta_0 \geq 1/2$, one readily verifies that $(1-\delta_0)/\ln(1/\delta_0) \geq 1/2$, and so $d = \Theta(\log(n))$.

If $\delta_0 < 1/2$ then $1 - \delta_0 \geq 1/2$ and $\ln(1/\delta_0) \leq 2\ln(\ln(n))$ by assumption (A2), and again for $n$ sufficiently large $d \geq 8$. □

*Proof of Property (P2).* $T_1^t$ is always equal to one of the sets $W^u$ (for some $u \leq t$), which has size at most $n+1$. Since $S_i^t$ is **middle**$(D)$ for a subsegment of $T_i^t$, $|S_i^t| \leq n/2$. □

*Proof of Property (P3).* We have $|S_{i-1}^t| \geq |T_i^t| \geq 2|S_i^t|$, since $S_i^t$ is contained in the middle of a subsegment of $T_i^t$. □

*Proof of Property (P4).* We have $|S_d^t| \geq \phi^{t-1}(S_d^t) \geq 1/\gamma$, by (25). □

*Proof of Property (P5).* This follows immediately by combining (18) and (19) and (I7). □

*Proof of Property (P6).* This follows from (24). □

*Proof of Property (P7).* Let $E$ be an epoch at level $i$. Here we are trying to lower bound $q_i^E$, which is the cost of all relocations done during epoch $E$. Let $s$ denote the start time, and $c$ denote the closing time, of epoch $E$. The busy segment $B^c$ includes a location outside of $S_i^E$ (this is the reason that the epoch closed at time $c$.) Without loss of generality let us say that $B^c$ includes a location that is to the left of $S_i^E$. Let $L$ be the left segment of $S_i^s - T_{i+1}^s$. The desired lower bound is an immediate consequence of the following four claims.

1. For each time $r \in E$, $T_{i+1}^r$ is a segment contained in $S^E$, $T_{i+1}^r = T_{i+1}^s \cup B_1 \cup \cdots \cup B_{r-1}$, and $B_r \cap T_{i+1}^r \neq \emptyset$.

2. Every key stored in $L$ at the start time $s$ of $E$ must move sometime during $E$.

3. The first step $t$ that a key $y$ stored in $L$ was moved during $E$ we have $y \in Q_i^t$. Thus $q_i^E$ is at least the number of keys that were stored in $L$ at step $s$.

4. The number of keys stored in $L$ at step $s$ is at least $|S_i^s|\rho^{s-1}(S_i^s)/8$.

For the first claim, it was noted in the remark after the description of the adversary that $T_{i+1}^r$ is a segment that intersects $B_{i+1}^r$. The fact that $T_{i+1}^r = T_{i+1}^s \cup B_s \cup \cdots \cup B_{r-1}$, comes from the definition of the adversary.

For the second claim, suppose for contradiction that $y$ is a key that is stored at location $j \in L$ at the end of time $s-1$ and does not move from $j$ throughout the epoch $E$. Then $j \notin T_{i+1}^s$ and $j \notin B_r$ for every $r \in E$. Then $j \notin T_{i+1}^s \cup B_s \cup \cdots \cup B_c$, which by the first claim, is a segment. This implies that $B_c$ contains no element to the left of $j$, contradicting that $B_c$ contains an element to the left of $L$.

For the third claim, consider the first step $t$ that $y$ was moved from location $j$. So $j \notin B^r$ for any $r \in [s,t)$, so is not in $T_{i+1}^t = T_{i+1}^s \cup B_s \cup \cdots \cup B_{r-1}$. Hence $j \in S_i^t - S_{i+1}^t$. Thus the relocation of $y$ is charged to level $i$ at step $t$.

For the fourth claim, $L$ is a subsegment of $S_i^t$ of size at least $|S_i^t|/4$. Since $S_i^t$ is $25\kappa$-lower balanced by (17), $\rho(L) \geq \rho(S_i^t)(1/4)^{25\kappa} \geq \rho(S_i^t)/2$, by (I6). □

This completes the proof of Lemma 11. □

### 6.4 Proof of Lemma 3

The hypotheses of Lemma 3 give us that $n$ is sufficiently large, $\delta_0 \leq (\ln(n)^{-2}, 1 - n^{-1/5})$, and mingap$(Y^0) \geq 1 + 12/\delta_0$. Then by Lemma 11 **Adversary**$(d, \kappa)$ satisfies (P1)-(P7) for $d, \kappa, \alpha, \gamma, \delta^*$ given by (10)-(14).

We apply Lemma 6 with these parameters. The denominator of (4) is $\Theta(\alpha d + \gamma\lambda + 1 - \delta_0)$. The settings given by (12) and (11) give $\alpha d \leq (1-\delta_0)/2$. The setting $\gamma = n^{-1/4}$ and $\lambda \leq 1$ and assumption (A2) give $\gamma\lambda \leq 1 - \delta_0$. So the denominator of (4) is $\Theta(1-\delta_0)$

For the numerator, the setting of $\delta^*$ gives $\delta^* \geq \delta_0/e$, the setting of $d$ gives $d^2 = \Theta((\ln(n))^2(1-\delta_0)^2/(\ln(1/\delta_0))^2)$. Simplifying the fraction gives:

$$\chi \geq \Theta\left(n \ln(n)^2 \frac{\lambda\delta_0(1-\delta_0)}{(\ln(1/\delta_0))^2}\right),$$

as required.

In the case that mingap$(Y^0) \geq 2^n$, $\lambda = 1$ and in the other case $\lambda = \Theta(\delta_0)$.

## 7 An upper bound for inserting a small number of items

In this section, we show an interesting upperbound on $\chi^n(m)$ for the case that $n$ is a polylogarithmic function of $m$.

**Theorem 14** *Let $m > 2^{16}$ and $k$ be an integer such that $k \leq 1/2\sqrt{\log m / \log \log m}$. Assume $n \leq \log(m)^{k/3}$. Then $\chi^n(m) \leq (2k-1)n$, i.e., there is an algorithm that loads $n$ keys into an array of size $m$ with amortized cost of $2k-1$ per key.*

**Proof:** We proceed by induction on $k$. To simplify the description we assume (without loss of generality) cells 1 and $m$ are initially loaded with keys $y_{\min}$ and $y_{\max}$ which are, respectively, lower and upper bounds on all keys. Set $Y^0 = \{y_{\min}, y_{\max}\}$.

At any time the array has certain occupied cells. A segment of cells whose leftmost and rightmost cells are occupied and all others are unoccupied is called an *open segment*; the keys in the leftmost and rightmost cells of the open segment $S$ are denoted $y_L(S)$ and $y_R(S)$ (we include the occupied end cells in the open segment for convenience in some calculations). The initial open segment has size $m$. The segment is said to be *usable* if $|S| \geq 3$ (which means there is at least one unoccupied cell). For any new key $y$ not stored in the array there is a unique open segment $S$ such that $y_L(S) < y < y_R(S)$; we say that $S$ is *compatible* with $y$. If key $y$ is assigned to an unoccupied cell in $S$ then the open segment $S$ is split into two open segments which overlap at the cell containing $y$; the sum of the sizes of these two segments is $|S| + 1$. A *middle cell* of $S$ is a cell such that the two segments obtained from $S$ each have size at least $|S|/2$. It's easy to check that every usable segment has a middle cell. More generally, it can be checked that given $q - 1$ items to be placed in an open interval $S$ that has at least $q - 1$ unoccupied spaces we can place them evenly so that each of the $q$ open segments produced has size at least $|S|/q$. (The worst case is $|S| = aq + 1$) for some integer $a$, and in this case each of the $q$ resulting subsegments has length $a + 1 \geq |S|/q$.)

We will define algorithms $A_k$ for $k \geq 1$. It will be obvious from the definitions that the cost per item loaded is at most $2k - 1$. The main technical question will be how many items $A_k$ can handle. Let us define $n_k(m)$ to be the maximum number of items that $A_k$ can handle in an interval of size at least $m$ (the argument $m$ need not be an integer). Our goal is to show that $n_k(m) \geq \lfloor \log(m)^{k/3} \rfloor$.

We now define algorithm $A_1$ for $k = 1$. For each successive key $y^t$ ($t \geq 1$), we identify the open segment $S$ compatible with $y^t$. If it is usable we store $y^t$ in the middle cell of the segment.

Let us analyze this algorithm. We never move any loaded key, so the cost per loaded key is 1. We want to lower bound the number of items that can be loaded. The size of the initial open segment is $m$, so after loading $t - 1$ items every open segment has size at least $m/2^{t-1}$ and we can handle $y^t$ if this is at least 3. Thus we can load $t$ items provided that $t - 1 \leq \log_2(m/3) + 1$. So $n_1(m) \geq \log_2(m/3) + 1$, which is at least $\log_2(m)^{1/3}$ for $m$ large enough.

For $k \geq 2$ we define $A_k$, which makes use of $A_{k-1}$. We initially load $q = \lfloor \log(m)^{(k-1)/3} \rfloor$ items using algorithm $A_{k-1}$, which by induction can be done at amortized cost of $2k - 3$ moves per item. We then move all of the items so that they are as evenly spaced as possible along the array which increases the amortized cost per item to $2k - 2$. Each of the resulting open segments has size at least $m/(q + 1)$. Let us define $s_j$ for $j \geq 0$ to be $2^{-j} m/(q + 1)$.

Next the algorithm works in *rounds*. Let old$^R$ denote the set of keys loaded prior to round $R$. The algorithm will ensure that the open segments defined by the locations of old$^R$ at the beginning of the phase all have size at least $s_{R-1}$. We have already seen that this holds when $R = 1$. Round $R$ will consist of two phases. During the first phase we will load $n_{k-1}(s_{R-1})$ new items without moving any items in old$^R$. During the second phase we move items (both old and new) to ensure that all open segments have size at least $s_R = s_{R-1}/2$.

During the first phase we refer to the open segments defined by the storage function at the beginning of the phase as *working segments*. We will run $A_{k-1}$ independently on each working segment. When a key arrives we assign it to the working segment it is compatible with, and load it into the working segment using $A_{k-1}$. Since each working segment has length at least $s^{R-1}$ and we load at most $n_{k-1}(s_{R-1})$ keys in all of the segments we are guaranteed that each of the independent copies of $A_{k-1}$ successfully load all of their assigned keys at amortized cost of $2k - 3$.

For the second phase we need to rearrange the elements to guarantee that the lower bound on working segment length decreases by at most a factor of 2. Classify keys as old or new depending on whether they were added in round $R$. Say that a segment $S$ is *useful* if its first and last cells contain old keys (useful segments are unions of one or more consecutive working segments), and define the excess of a useful segment to the number of old keys in it minus the number of new keys. We need to identify a collection of disjoint segments each having excess exactly one that cover all of the new keys. To see that such a collection exists, first note that the entire array is a useful segment with positive excess. Now choose a collection $\mathcal{S}$ of disjoint useful segments that together cover all new keys, such that $\mathcal{S}$ is as large as possible, and subject to this, the sum of the sizes of segments in $\mathcal{S}$ is as small as possible. We claim that each $S \in \mathcal{S}$ has excess exactly one. Consider such an $S$ and assume that its excess $m$ is greater than 1. Let $j_0, j_1, \ldots, j_t$ be the index of the cells of $S$ that store old elements. If there are no new elements between $j_0$ and $j_1$ then we can shrink $S$ to start at $j_1$ contradicting the minimality of $\sum_{T \in \mathcal{S}} |T|$. Thus the segment $[j_0, j_1]$ has excess at most 1. Let $j_r$ be the largest index such that $[j_0, j_r]$ has excess at most 1. Then $[j_0, j_{r+1}]$ has excess greater than 1, which implies that $[j_0, j_r]$ has excess exactly 1 and there are no new keys in $[j_r, j_{r+1}]$. But then we can split $S$ into $[j_0, j_r]$ and $[j_{r+1}, j_t]$ each of which has positive excess, and this contradicts the maximality of $\mathcal{S}$. So $\mathcal{S}$ has the desired properties.

Now within each segment $S \in \mathcal{S}$ we redistribute the keys (both old and new) that are internal to $S$ uniformly within $S$. If $S$ had $u - 1$ internal old keys then it was originally split into $u$ working segments each of size at least $s_{R-1}$. We now have $2u - 1$ internal keys which will split $S$ into $2u$ working segments (that overlap at their endpoints) in the next round and after redistributing them all of them will have size at least $s_{R-1}/2 = s_R$.

The total work done to accomplish phase 2 is $2u - 1$ which is less than twice the number of new keys in the segment. Summing over all segments in $\mathcal{S}$ gives an additional amortized cost of 2 per key. Thus the total amortized cost of $A_k$ is at most $2k - 1$ per item.

It remains to bound the number of items that can be handled by $A_k$.

Let $r$ denotes the number of rounds. We have $s_1 = m/q > \frac{m}{\log(m)^{(k-1)/3}}$. Then the number of items inserted during all rounds is

$$\sum_{i=1}^{r} \log_2(s_i)^{(k-1)/3} \geq \sum_{i=1}^{r} \log_2\left(\frac{s_1}{2^i}\right)^{(k-1)/3} =$$

$$= \sum_{i=1}^{r} (\log_2 s_1 - i)^{(k-1)/3} \geq r(\log_2 s_1 - r)^{(k-1)/3}$$

Let us chose $r = \sqrt{\log m}$ and we obtain

$$\sqrt{\log_2 m} \left( \log_2 \frac{m}{3 \log_2(m)^{(k-1)/3}} - \sqrt{\log_2 m} \right)^{(k-1)/3} =$$

$$\left( \log_2(m)^{(2k+1)/6} \right) \left( 1 - \frac{\sqrt{\log_2 m}}{\log_2 m} - \frac{k-1}{3} \frac{\log_2 \log_2 m}{\log_2 m} \right)^{(k-1)/3}$$

Using the fact that $k < 1/2\sqrt{\log_2 m / \log_2 \log_2 m}$ we obtain the lower bound for this expression

$$\left( \log_2(m)^{(2k+1)/6} \right) \left( 1 - \sqrt{\frac{\log_2 \log_2 m}{\log_2 m}} \right)^{\frac{1}{6} \cdot \sqrt{\frac{\log_2 m}{\log_2 \log_2 m}}} \geq \left( \log_2(m)^{(2k+1)/6} \right) \left( \frac{1}{2e} \right)^{\frac{1}{6}}$$

which is for $m > 2^{16}$ greater than $\log_2(m)^{2k/6}$. Therefore during all rounds of $A_k$, $\log_2(m)^{k/3}$ keys are loaded with an amortized cost $2k - 1$ per insertion.

□

## References

[BCD+02] Michael A. Bender, Richard Cole, Erik D. Demaine, Martin Farach-Colton, and Jack Zito. Two simplified algorithms for maintaining order in a list. In *Proceedings of the 10th Annual European Symposium on Algorithms*, ESA '02, pages 152–164, 2002.

[BDFC05] Michael A. Bender, Erik D. Demaine, and Martin Farach-Colton. Cache-oblivious B-trees. *SIAM J. Comput.*, 35:341–358, 2005.

[BDIW04] Michael A. Bender, Ziyang Duan, John Iacono, and Jing Wu. A locality-preserving cache-oblivious dynamic dictionary. *Journal of Algorithms*, 53(2):115 – 136, 2004.

[BFJ02] Gerth Stølting Brodal, Rolf Fagerberg, and Riko Jacob. Cache oblivious search trees via binary trees of small height. In *Proceedings of the thirteenth annual ACM-SIAM symposium on Discrete algorithms*, SODA '02, pages 39–48. Society for Industrial and Applied Mathematics, 2002.

[BS07] Richard S. Bird and Stefan Sadnicki. Minimal on-line labelling. *Inf. Process. Lett.*, 101:41–45, January 2007.

[DSZ05a] Paul F. Dietz, Joel I. Seiferas, and Ju Zhang. Lower bounds for smooth list labeling. *Manuscript*, 2005.

[DSZ05b] Paul F. Dietz, Joel I. Seiferas, and Ju Zhang. A tight lower bound for online monotonic list labeling. *SIAM J. Discret. Math.*, 18:626–637, March 2005.

[IKR81] Alon Itai, Alan G. Konheim, and Michael Rodeh. A sparse table implementation of priority queues. In *Proceedings of the 8th Colloquium on Automata, Languages and Programming*, pages 417–431, 1981.

[Wil92] Dan E. Willard. A density control algorithm for doing insertions and deletions in a sequentially ordered file in a good worst-case time. *Inf. Comput.*, 97:150–204, April 1992.

[Zha93] Ju Zhang. *Density Control and On-Line Labeling Problems*. PhD thesis, University of Rochester, Rochester, NY, USA, 1993.